\documentstyle[prl,aps,multicol,epsf]{revtex}

\begin{document}
\draft
\title{Statistical Models of the Polaronic \\
Phase Transition in Manganites}
\author{Igor F.Lyuksyutov$^{a,*}$ and Valery Pokrovsky $^{a,b}$}
\address{(a) Department of Physics, Texas A\&M University,\\
College Station, TX 77843-4242 \\
(b) Landau Institute for Theoretical Physics, Moscow, Russia}
\date{\today}
\maketitle

\begin{abstract}
We propose two statistical models for description 
the metal-insulator phase transition coupled 
with paramagnetic-ferromagnetic
phase transition in manganites of the type $La_{1-x}Sr_xMnO_3$.
The first one based on the competition of small 
polarons and delocalized carriers.
In the second one the conductivity appears 
as a result of overlapping of large
polarons. We compare our models with the experimental data. 
\end{abstract}


\begin{multicols}{2}
\narrowtext
     
The interplay of the metal-insulator (M-I) and ferromagnetic-paramagnetic
(FM-PM) transitions in mixed metal oxides is one of the most important and
least understood topics in material physics. 
The early explanation of the combined metal-insulator (M-I) and
ferromagnetic-paramagnetic (FM-PM) transitions was based on the
Double-Exchange (DE) mechanism \cite{zen}.
In this mechanism the exchange interaction 
between spins is provided by conduction
electrons. The electron tunneling amplitude from one site to another depends
strongly on orientation of spins on these sites. Therefore, magnetic phase
transition causes simultaneous drop in resistivity.

Recently, Millis {\it et al}. \cite{mill1},\cite{mill2} have shown that the Double
Exchange itself does not explain the transition 
from the conducting to insulating phase.
They argued that Jahn-Teller distortions are strong enough and lead to
formation of polarons. The high temperature phase in their picture contains
presumably polaronic carriers and is insulating \cite{mill2}. In the low
temperature phase delocalized carriers appear together with the
magnetization. This idea is supported by experimental observations of large
Jahn-Teller distortions \cite{booth}. It was confirmed experimentally that
small polarons are responsible for transport at high temperature \cite{hundley}.
For reviews on the experimental state of art see \cite{ram}.

Quantitative description of the phenomenon was based on a phenomenological
Hamiltonian proposed by Millis, Schraiman and Mueller (MSM) \cite{mill2} which 
describes electrons coupled via elastic
deformations and interacting with localized spins 

\begin{eqnarray}
\nonumber
{\cal H}&=&\sum_{{\bf r,a}}t_{{\bf r,r+a}}\psi^{\dagger}({\bf r})\psi ({\bf r+a})
-\sum_{\bf r}J_{H}{\bf S}{_{{\bf r}}}\cdot\psi^{\dagger}({\bf r}){\bf \sigma }\psi ({\bf r}) 
\\ &+&\sum_{\bf r}\left({\kappa\over 2}{\bf Q}{_{{\bf r}}^{2}}
-g{\bf Q}{_{{\bf r}}}\cdot \psi^{\dagger}({\bf r}){\bf \tau }\psi({\bf r})
 - \mu\psi^{\dagger}({\bf r})\psi 
({\bf r})
\right) 
\label{h}
\end{eqnarray}

Here $t_{{\bf r,r+a}}$ is the nearest neighbor tunneling amplitude, 
 $\psi ({\bf r}), \psi^{\dagger} ({\bf r})$ are fermion operators, ${\bf Q}_{{\bf r}}$ 
are local Jahn-Teller-displacements, $\sigma$ and $\tau$ are
spin and pseudo-spin Pauli matrices, $J_{H}$ is the Hund energy, $g$ is the
electron-phonon coupling constant and $\kappa $ is the elastic 
coefficient. $J_{H}$ is assumed to be the largest energy 
scale in the problem. In this
approximation electron spins  follow adiabatically the direction of
localized spins on a site \cite{mill2}. The authors \cite{mill2} treated
their model Hamiltonian in the mean field approximation.
For the number of carriers
per site $n=1$ they have found an insulator phase with strong 
Jahn-Teller distortion. For $n<1$ the high temperature phase in the
frameworks of this approximation remains conducting, though the conductivity
can decrease rapidly above transition temperature. In particular, no remarkable
magneto-resistance have been found at experimentally optimal value $n=0.3$.
These results allow a following interpretation. The fermion variables can be
eliminated explicitly from the partition 
function based on the Hamiltonian (\ref{h}).
After elimination we arrive at an effective 
free energy for variables $\bf Q$ and
$\bf S$:
\begin{equation}
\frac{F}{T}=- Tr\ln{\left(1+\exp{(\hat{t}-\hat{J}-\hat{Q}+\hat{\mu})}
\right)}-{\kappa \over 2}\sum_{\bf r}{Q_{\bf r}^2}
\label{heff}
\end{equation}
where operator $\hat{\mu}$ is diagonal $4N\times 4N$ matrix in 
variables ${\bf r},\sigma, \tau$, whereas operator $\hat{t}$ 
is off-diagonal only with respect to
variable $\bf r$ with non-zero matrix 
elements $t\sum_{\bf a}\delta_{\bf r,r^{\prime}+a}$; other two 
operators $\hat{\bf J}$ and $\hat{\bf Q}$ are off-diagonal
with respect to variables $\sigma$ and $\tau$ with matrix elements 
 $J_H{\bf S}_{\bf r}\cdot{\bf \sigma}_{\sigma ,\sigma^{\prime}}$ and
 $g{\bf Q}_{\bf r}\cdot {\bf \tau}_{\tau ,\tau^{\prime}}$, respectively.
The mean-field (Landau) counterpart of the free energy (\ref{heff}) reads:
\begin{equation}
F_L\,=\,a{\bf S}^2+b({\bf S}^2)^2+A{\bf Q}^2+B({\bf Q}^2)^2+c{\bf S}^2{\bf Q}^2.
\label{landau}
\end{equation}
This is well-known problem of two interacting order parameters 
(see, e.g. \cite{PP}). Its typical phase diagram contains 
4 phases with $\bf S$ and $\bf Q$ either equal to
zero or non-zero. It is tempting to identify phases with ${\bf Q}\neq 0$ as
insulating and those with ${\bf Q}=0$ as metallic. 
Then 4 mentioned phases would be
paramagnetic metal (PM), paramagnetic insulator (PI), 
ferromagnetic metal (FM) and 
ferromagnetic insulator (FI). However, this model 
being applied for description of 
manganites with CMR
has two substantial failures: i) its high-temperature 
phase is metallic, whereas in
real manganites it is insulating; ii) the transition from 
the PI to the FM phase
proceeds at one point only, whereas in real systems it proceeds along a line.
Therefore, the mean-field approximation applied 
by MSM from our point of view is
insufficient to explain the real phase diagram in manganites.  
It happens because any
solution with the homogeneous displacement field ${\bf Q}$ does not describe
formation of a localized objects - polarons. The fact that MSM obtained large
magneto-resistance at $n\approx 1$ is readily explained as a result of splitting of
the double-degenerate upper level by strong 
Jahn-Teller effect. But this effect does
not work properly at smaller values of $n$, 
since the lower band is not fully occupied. 

In this article we propose
simple statistical models in which polaron states are incorporated
phenomenologically. These models have 
the two most important phases found in
the experiment: paramagnetic insulator 
(PI) and ferromagnetic metal (FM).
In the first model which we call Small Polaron (SP) Model 
we start with the free energy written in the form:

\begin{equation}
{\cal F}=(-(1+m^{2})+{\frac{\beta }{2}}m^{4})tn_c+Tm^{2}
+{\cal F}_{LG}
\label{f}
\end{equation}
where ${\cal F}_{LG}= -n_{p}\varepsilon +T(n_c\ln n_{c} +n_{p}\ln n_{p}) $ is the small 
polaron lattice gas free energy.
$n_c$ is the density of conductance electrons, $n_{p}=n-n_c$ is
the density of localized electrons (small polarons), $n$ is the total
electron density; $t$ is the bandwidth, $-\varepsilon $ is the small polaron
energy, $m$ is the magnetization; the constant $\beta $ allows to vary the
saturation magnetization. In the insulating (polaron) phase $m=0$.
Minimizing free energy Eq.\ref{f} with respect to $m$ one finds equation 
for $x=n_c/n$:  $x\tilde{\Delta}+\frac{T^{2}}{2n^{2}x^{2}}+T\ln {\frac{x}{1-x}}=0 $.
Depending on parameters, 
this equation has one, two or no roots. For $\tilde{\Delta}>0$ 
and $T=0$ this equation  has no roots. It means that the 
polaronic phase is stable. For $\tilde{\Delta}<0$ the FM 
phase is stable at $T=0$. Since the polaronic phase
always wins at large temperature, it implies that for $\tilde{\Delta}<0$ the
PI-FM phase transition takes place at any value of $n.$. The phase diagram
was calculated numerically.
For $\beta <1/(n\ln 2)$ the transition
temperature grows monotonously with $n$. If $\beta >1/(n\ln 2)$ the
ferromagnetic phase exists not at any $n$, but only at the carrier density
exceeding a critical value $n_{c}<n$. An interesting feature of this model
is the possibility of re-entrant polaronic phase at low temperature. This
peculiarity may be related to the reentrant behavior of resistivity found in
the experiment \cite{urushibara}.  The energy parameters of the 
model $\varepsilon $ and $t$ are in the range $1eV,$ 
but numerical values of the
transition temperature are relatively 
small: $ant<T_{c}<bnt$, where $a\approx 0.01,b\approx 0.3$.

\begin{figure}
\begin{center}
\epsfxsize=.8\linewidth
\epsfbox{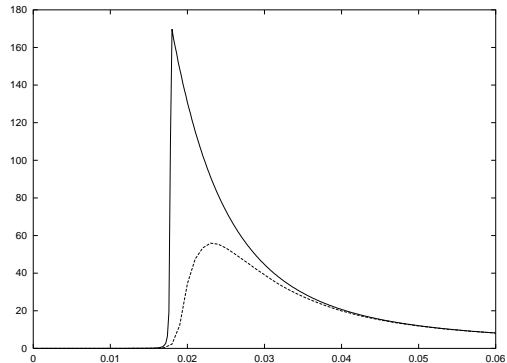}
\end{center}
\caption{Resistivity $\rho  $(arbitrary units) vs  $T/t$   without
and with (dashed line) effective magnetic field $h_0$ (see text).
\label{rho}}
\end{figure}

Though the SP model discussed above displays PI and FM phases and the
colossal magneto-resistance, some its features seem unrealistic. It implies
the first order phase transition with rather large 
discontinuities of the values $m, n_c$ and entropy
for a reasonable set of parameters, 
whereas there is no experimental
indication of remarkable hysteresis near the transition line. 
These discontinuities may be smeared out by a small number 
of impurities or other defects. We simulated the action 
of defects by averaging of the resulting values $m(T/t,\varepsilon ), n_c(T/t,\varepsilon )$ 
over a small range ($3\div 4\%$) 
of the parameter $\varepsilon$. The results 
are shown in Fig.\ref{rho} for zero and small finite magnetic field.
In this calculation we identified
 $n_c^{-1}$ with the resistance. In reality they 
differ by a factor (mobility) which
varies slowly, at least near the transition line. 
Another unrealistic feature of the
SP model is too large scale of magnetic fields 
(this feature is shared with the
MFA calculations of the MSM work). If one accepts the intersite
tunneling constant equal to $0.6 eV$ \cite{pickett}, 
the essential magnetic field at room temperature 
is about 10-100 times more than experimental values
(field $h_0$ in Fig.\ref{rho} corresponds to $\sim 50 T$ for $t = 0.6 eV$). 
Our second model was inspired by the neutron scattering 
experiments \cite{hennion}, \cite{louca} which
showed the occurrence of large magnetic polarons (LP) coexisting with the
small polarons (SP). This phenomenon lies beyond the scope of the simple SP
model. Below we discuss an alternative mechanism of the PI-FM phase
transition based on the LP.

We employ Varma theory \cite{varma} modified to incorporate the Jahn-Teller
effect. The free energy of an individual carrier localized in a sphere of
radius $R$ reads:
\begin{equation}
F(R)=-W+\frac{t}{R^{2}}-CTR^{3}-\frac{\varepsilon_0}{R^{3}}
\label{fl}
\end{equation}
Here $W\propto t$ is the half-width of the conductivity band, $R$ is
measured in the lattice constants, $t$ and $\varepsilon_0\sim \varepsilon $ have the same
meaning as before; C is a numerical constant which can be roughly estimated 
as  $(4/3)\pi\ln{4}=5.8$. The latter figure is probably 
overestimated since it is found
for the case of full saturation of magnetization 
in large polaron which is scarcely
realized. Further we assume $C$ to be of the order of one. 
The first three terms in the r.-h. s. of eqn. \ref{fl}
were discussed in \cite{varma}. The last term describes the polaronic
effect. It contributes a term proportional to ${|\Psi ({\bf R})|}^{4}$ 
to the energy density where $\Psi ({\bf R})$ 
is the electron wave-function (see, for example 
\cite{emin}). The contribution to the total energy  is inverse
proportional to the volume of the localized state. Exactly as in the work 
\cite{varma}, the localization on large scale of $R$ originates from the
entropy term $TR^{3}$. The Jahn-Teller term can be 
neglected if $t$ and $\varepsilon_0$ have the same order 
of magnitude and $T$ is small. Minimizing $F(R)$, 
we find the radius of the stable or metastable 
polaron $R_{p}\approx(2t/3T)^{1/5}$. This result shows that 
the linear size of the LP changes
very slowly with temperature. At reasonable values of $t\sim 0.6eV$ 
\cite{pickett} and $T\sim 0.01\div 0.03eV$ it does not exceed $2.5$.
Thus, the large polaron is not so large. Still it contains between 30 and 60
sites and its magnetic moment is large.

The ratio of numbers of the LP and SP is given 
by the Boltzmann factor $\exp[(\varepsilon -F(R_{p}))/T]$ 
where $\varepsilon $ is, as above, the energy of the small polaron. 
We introduce the value $\Delta _{p}=\varepsilon +W$.
If $\Delta _{p}$ is positive, this ratio tends to 
a small number $\exp(-R_{p}^{3})$ at high temperature. 
At some low enough temperature $T_{p}$
the LP density becomes equal to the inverse volume of the LP. This
temperature can be found from equation:
 $\frac{\Delta _{p}}{T_{p}}-1.96\left( \frac{t}{T_{p}}\right) ^{3/5}+\ln\left( nV_{p}\right) =0$
where $V_{p}=(4/3)\pi R_{p}^{3}$ is the volume of the LP. At this (or
slightly higher) temperature large polarons overlap forming an infinite
cluster. The percolation leads to a rapid decrease of resistivity. Indeed,
the magnetization inside the cluster is homogeneous. The condition at which
the spontaneous magnetization appears was shown to be $tn_c>$ $T$. The
concentration of delocalized carriers $n_c$ 
in the cluster is equal to $V_{p}^{-1}$. Thus, $tn_c/T_{p}\approx (t/T_{p})^{2/5}>1$.

Normally the percolation transition proceeds continuously. 
The Coulomb repulsion prevents formation 
of macroscopic magnetic droplets before
percolation. The Coulomb energy $e^{2}/\epsilon R$ must be compared with the
kinetic energy $t/R^{2}$. Even for static dielectric constant $\epsilon =25$
the Coulomb energy prevails. We expect that the screening is even weaker at
small distances.

It is important to note that the existence of magnetization inside the
cluster does not guarantee the total non-zero magnetization. The percolation
and the spontaneous magnetization appearance proceed simultaneously in 2
dimensions. The reason is that two infinite clusters for different spin
orientations would unavoidably intersect. In 3 dimensions they can
interpenetrates without intersection as long as their branches are thin
enough. The spontaneous magnetization appears when the cross section of each
infinite cluster increases. This fact 
was established for Ising model \cite{dotsenko}, 
but its geometrical origin is quite general. In the frameworks
of the LP model the Mott transition proceeds, generally speaking, at a
temperature higher than the Curie point.

An approximate solution of equation for $T_p$ is: $T_{p}\simeq t(\Delta_{p}/2t)^{5/2}$. 
At $\Delta _{p}\approx 0.5t$ we find $T_{p}\simeq 0.03t\approx 200K$. 
This estimate looks reasonable. The logarithmic 
term in  equation for $T_p$ provides a weak dependence of $T_{p}$ 
on the dopant concentration $n$.

\begin{figure}
\begin{center}
\epsfxsize=.8\linewidth
\epsfbox{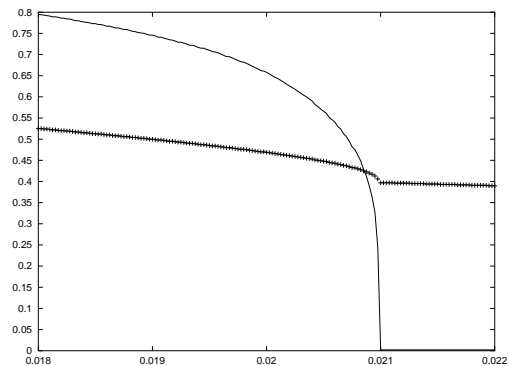}
\end{center}
\caption{Magnetization (line) and large polaron concentration ($+$) vs  $T/W$
for the set of parameters $\bar{J}=.2W$ $\bar{K}=0.6 W$.
\label{hei}}
\end{figure}

Statistics of LP can be constructed via a kind of 
the lattice model with the sites
either occupied by a LP or empty (the elementary 
cell of this auxiliary lattice has
the volume $V_p$). Introducing occupation numbers $\tau_{\bf r}=0,1$ and polaron 
spins ${\bf S}_{\bf r}=S{\bf n}_{\bf r}$ on sites, the proposed lattice Hamiltonian
reads:
\begin{eqnarray}
\nonumber
{\cal H}_{lp}&=&-\sum_{\bf r,a}J{\bf n}_{\bf r}\cdot{\bf n}_{\bf r+a}
\tau_{\bf r}\tau_{\bf r+a}\\
&+&\sum_{\bf r,a}K\tau_{\bf r}\tau_{\bf r+a}
-\sum_{\bf r}({\bf h}\cdot{\bf n}_{\bf r}\tau_{\bf r}+\mu\tau_{\bf r}),
\label{lattice}
\end{eqnarray}
where $J$ is the effective exchange constant, $h=g\mu_BHS$ 
is an effective magnetic
field for polarons, $K$ stays for the Coulomb interaction 
between neighboring polarons
and $\mu=W-CW^{3/5}T^{2/5}$ is the polaron free energy. 
The mean-field counterpart of
the Hamiltonian (\ref{lattice}) reads:
\begin{equation}
F_{lp}=-\bar{J}m^2x^2-T{\cal S}_H(m)-hmx-{\cal F}_I(x)
\label{LPL}
\end{equation}
where $m=|\langle{\bf n}_{\bf r}\rangle|$ and $x=\langle\tau_{\bf r}\rangle$,
 $\bar{J}=zJ$, $\bar{K}=zK$, $z$ is coordination number.
 ${\cal F}_I(x)=\bar{K}x^2 - T [x\ln{x}+(1-x)\ln{(1-x)}]-\mu x.$
 ${\cal S}_H(m)$ is the spin 
entropy per site for the Heisenberg system. 
For large $S$ we are interested in, 
 ${\cal S}_H(m)$ is determined by a following equations:
 ${\cal S}_H(m)\,=\, \ln{2S}\,-\,\ln{\left(1-(1-m)\zeta (m)\right)}\,-\,(m+1)\zeta (m),$
where $\zeta (m)$ is determined by equation:
 $\coth\zeta - 1/\zeta\,=\,m$.
We calculated magnetization by minimization 
of free energy (\ref{LPL}) and resistance
employing the effective medium approximation \cite{effmedia}. 
The results for magnetization and large polaron density
are shown in Fig.\ref{hei}.  
Despite of visible similarity 
of the LP mean-field model to corresponding
SP model, the former displays the second 
order transition in a broad range of
parameters. The resistivity show magnetic field
dependence qualitatively similar to that 
in Fig.\ref{rho}. However, due to large 
value of $S$ for a polaron, the scale of
magnetic field displaying the CMR is about 
100 times lower than for the SP model and,
thus, it fits satisfactory the experimental data.
Transition temperature in Fig.\ref{hei} is in qualitative agreement 
with estimate from the simple large polaron model of
Eq.\ref{fl}. 
The phase diagram of the model Eq.\ref{LPL} displays
region of the second order phase transitions 
for  $0.5 W \leq \bar{K}\leq 1.5 W$ and for  $\bar{J}\leq 0.2\div 0.3 W$ 
with $T_p \propto \bar{J} $.
The Mean Field Approximation (MFA) is insensitive to percolation 
threshold near which the phase transition occur. 
However, one can estimate  correction to the MFA phase diagram
by using results known
from dilute magnetic models \cite{stinch}.
Corresponding corrections decrease the transition
temperature in the region with LP concentration larger than
percolation threshold by $\approx 10\% \div  20\% $. 
 More detailed description of the
phase diagram will be given elsewhere.

In conclusion, we proposed statistical 
models explicitly incorporating
polaron formation in manganites. One of them 
(the SP model) displays first order
phase transition from
paramagnetic state with localized electrons 
to the ferromagnetic state with
metallic conductivity in perovskites, 
the colossal magneto-resistance and
the re-entrant resistivity in a range of 
variables. The second LP model is
based on percolation of large polarons. 
It predicts a continuous isolator-conductor transition 
at a temperature higher than the magnetic
transition. Experimental data on neutron 
scattering, the magnetic and resistive
measurements near the transition line allow 
to give some preference to the LP model,
though the further experimental and theoretical 
work is necessary. When this work 
was completed and presented at the Argonne Symposium on
Poorly Conducting Oxides (Argonne, June 1998), 
R.H. Heffner kindly informed us that he
and his colleagues were driven to a similar conclusion 
on the existence of dynamic
conducting clusters treating results of their 
experiments on muon spin relaxation in $La_{1-x}Ca_xMnO_3$ \cite{heffner}.

This work was partly supported by the grants 
NSF DMR-97-05182 and THECB ARP 010366-003.
It is a pleasure to acknowledge discussions with 
D.G. Naugle, R.H. Heffner, J. Lynn and A.J. Millis.

\end{multicols}


\begin{references}
\bibitem[*]{byline}  Also at Institute of Physics, 252028 Kiev, Ukraine
\bibitem{zen}  C. Zener, Phys. Rev.. {\bf 82}, 403 (1951);
P.W. Anderson and H. Hasegawa, Phys. Rev. {\bf 100}, 675
(1955);
P.G. De Gennes, Phys.Rev. {\bf 118}, 141 (1960).
\bibitem{mill1}  A.J. Millis, J. Littlewood, and B.I. Shraiman, Phys. Rev.
Lett. {\bf 74}, 5144 (1995).
\bibitem{mill2}  A.J. Millis, B.I. Shraiman, and R. Mueller, Phys. Rev.
Lett. {\bf 77}, 175 (1996).
\bibitem{booth}  C.H. Booth, F. Bridges, G.H. Kwei, J.M. Lawrence, A.L.
Cornelius, and J.J. Neumeier, Phys. Rev. Lett. {\bf 80}, 853 (1998).
\bibitem{hundley}  M.F. Hundley, M. Hawley, R.H. Heffner, Q.X. Jia, J.J. Neumeier, 
J. Tesmer, J.D. Thomson, and X.D. Wu, Appl. Phys. Lett. {\bf 67}, 860 (1995).
M. Jaime {\it et al.}, Appl. Phys. Lett. {\bf 68}, 1576 (1996).
\bibitem{ram}  A.P. Ramirez, J. Phys., Cond. Matt. {\bf 9}, 8171 (1997).
\bibitem{PP} A.Z. Patashinskii and V.L. Pokrovskii, Fluctuation Theory of Phase
Transitions, Pergamon, London, 1979.
\bibitem{urushibara}  A. Urushibara, Y. Morimoto, T. Arima, AA. Asamitsu, G.
Kido, Y. Tokura, Phys. Rev. B {\bf 51}, 14103 (1995).
\bibitem{hennion}  F. Moussa, M. Hennion, J. Rodriguez-Carvajal, J. Moudden,
L. Pinsard and A. Revcolevschi, Phys. Rev. B {\bf 56}, R497 (1997);
Journ. Magn. Magn. Mat, {\bf 177}, 858 (1998).
\bibitem{louca}  D. Louca , T. Egami, E.L. Brosha, H. Roder, and A.R.
Bishop, Phys. Rev. B {\bf 56}, 8475 (1997); T. Egami, D. Louca, R.J. McQueeney,
J.Superconductivity {\bf 10}, 323 (1997).
\bibitem{varma}  C.M. Varma Phys.Rev.B {\bf 54}, 7328 (1996).
\bibitem{emin}  D. Emin, in {\it Localization and Metal-Insulator Transition}
p. 323, edited by H. Fritzsche and D. Adler, Plenum Press, New-York, 1985;  
\bibitem{pickett} W.E. Pickett and D.J. Singh, Phys. Rev. B {\bf 53}, 1146 (1995).
\bibitem{dotsenko} V.S. Dotsenko, M. Picco, P. Windey, G. Harris, E. Martinec, and 
E. Marinary, Nucl. Phys. B {\bf 448}, 577 (1995).
\bibitem{effmedia} J.P.Clerc, G.Giraud, J.M.Laugier, and J.M.Luck, Adv.Phys. 
{\bf 39}, 191 (1990).
\bibitem{stinch}  R.B.Stinchcombe, in {\it Phase Transition and Critical Phenomena}
v.7 p.150, edited by C.Domb and J.L.Lebowitz, 
Academic Press, London, New-York, 1983;  
\bibitem{heffner} R.H. Heffner, L.P. Le, M.F. Hundley, J.J. Neumeier, G.M. Luke, 
K. Kojima, B. Nachumi, Y.J. Uemura, D.E. MacLaughlin, 
S.W. Cheong, Phys. Rev. Lett.{\bf 77}, 1869 (1996); 
R.H. Heffner, M.F. Hundley, and C.H. Booth, in Mat. Res. Soc.
Proc. {\bf 494} (1998).
\end{references}
\end{document}